\documentclass [12pt,a4paper] {article}

\usepackage {amsmath}
\usepackage {amssymb}
\usepackage {graphicx}

\begin {document}

\title {The Planck length and the constancy of the speed of light in five
dimensional space parametrized with two time coordinates}
\author {Christoph K\"ohn}
\date {Technical University of Denmark, National Space Institute (DTU
Space), Elektrovej 328, 2800 Kgs Lyngby, Denmark \\[2ex]
\today}

\maketitle

\begin {abstract}
In relativity and quantum field theory, the vacuum speed of light is
assumed to be constant; the range of validity of general relativity
is determined by the Planck length. However, there has been no convincing theory
explaining the constancy of the light speed. In this paper, we assume a
five dimensional spacetime with three spatial dimensions and two local time
coordinates giving us a hint about the constancy of the speed of light. By decomposing the
five dimensional spacetime vector into four-dimensional vectors for each time dimension
and by minimizing the resulting action, for a certain class of additional
time dimensions, we
observe the existence of a minimal length scale, which we identify as the
Planck scale. We derive an expression for the speed of light as a function
of space and time and observe the constancy of the vacuum speed of light
in the observable universe.
\end {abstract}

\section {Introduction} \label{intro.sec}
Since Maxwell's theory of classical electrodynamics \cite{maxwell_1864}, it
has been known that all electromagnetic waves travel with the speed of light.
In 1864 and 1881 experiments performed by Michelson and Morley
\cite{michelson_1881,michelson_1887} gave a hint that electromagnetic waves travel equally fast in all
inertial systems. This result was confirmed
by many more experiments \cite{essen_1955, trimmer_1973, wolf_2003,
herrmann_2009,nagel_2015} proofing the constancy of the speed of light
within the validity of the laboratory setups.

When Albert Einstein derived special relativity \cite{einstein_1905}, he
postulated that the speed of light be constant, and he used this
assumption as a key ingredient for special relativity. Consequently, he derived that the
speed of light is the upper velocity limit, be it for particles or for
information. As such, the limitation of velocities ensures causality; vice
versa, information sent with velocities above the speed of light could
eventually harm causality. 

However, up to now, there have not been
sufficiently convincing explanations why the speed of light is constant and why it has the
value which it has. Loop quantum gravity, for example, dictates that
the velocity of a photon is not defined to be constant, but has a value depending on its frequency \cite{camelia_1998}.

Indeed, there have been suggestions that its value might
vary with the age of the universe and that it might not
have been constant in the early stages of the universe. Albrecht and Magueijo
\cite{albrecht_1999} show that the cosmological evolution equations
together with a variable speed of light
might solve the horizon, flatness and cosmological
constant problem and together with cosmological perturbations the homogeneity and
isotropy problem. Deriglazov and Ram{\'i}rez \cite {deriglazov_2015,deriglazov_2016,
deriglazov_2017} observed a discrepancy between the speed
of light and the critical speed in the theories of spinning particles on
curved and electromagnetic backgrounds. Additionally they noticed that the
constancy of the speed of light is closely related with the self-consistent
definition of the three-acceleration in general relativity.

The necessity of deriving a theory for quantum gravity
resides from the problem that general relativity loses its validity
at small length scales \cite{wheeler_1955,klein_1956} and that, for
example, quantum electrodynamics stops being a self-consistent theory
if gravitational effects \cite{landau_1954} are added to the theory. The length scale
of these effects is in the order of
approximately $10^{-35}$ m and was already introduced by Planck in 1899
\cite{planck_1899} after the discovery of the Planck's constant when he
realized that he could
derive a unit system depending on the gravitational constant, speed of
light, Planck's constant, Boltzmann's constant and Coulomb's constant only.

Recently Faizal \cite {faizal_2014} and Pramanik et al. \cite {pramanik_2015} used the Planck length to
investigate the deformed Heisenberg algebra. Based on this algebra Faizal investigated the
deformation of the Wheeler-DeWitt Equation and thus showed that
the big bang singularity gets obsolete. Additionally, Faizal et al. \cite{faizal_2016} adopted the idea of a minimum
length scale or equivalently of a minimum time scale and showed that this
leads to corrections to all quantum mechanical systems by the deformed
Heisenberg algebra and thus to a discrete spectrum for time.

One of the attempts to unify general relativity with the quantum description
of the microscopic cosmos has led to string theory
\cite{bardakci_1969,nambu_1970a,green_1987,luest_1989}. One of its
main features is the existence of 10 or 26 space dimensions which is larger
than the commonly experienced three space dimensions \cite{witten_1995}. These extra dimensions are assumed to be compactified and are
consequently too small to be observed.

On the contrary, there have not been many approaches to a second or more time
dimensions. Tegmark \cite{tegmark_1997} summarized that a universe with a large second time dimension
cannot contain observers, like us humans, because of the lack of causality.
Hence, as for the additional space dimensions in string theory, extra time
dimensions need to be compactified in case of their existence.

Multitemporal spacetime dimensions have for example been discussed by Bars
and Kounnas \cite{bars_1997a,bars_1997b}; they consider two
time dimensions and construct actions for interacting $p$-branes
within two dimensions. They show that after a phase transition the additional
time dimension becomes part of the compactified universe. Additionally, they present a new
Kaluza-Klein like dimensional reduction mechanism and propose
an action for a string in two time dimensions. Due to new
gauge symmetries, they observe that quantum constraints are consistent only in
spacetime dimensions with signature (25,2) or (26,2) for a bosonic string or
(9,2) or (10,2) in the supersymmetric case.

Chen \cite{chen_2008} interprets two extra
time dimensions as quantum hidden variables and shows that non-local
properties of quantum physics or that the de Broglie wave length are natural consequences of the
existence of two additional time dimensions.

We here now suggest the existence of a compactified second time dimension and
subsequently derive the existence of a smallest length scale, i.e. the Planck
length, and explain the constancy of the speed of light in our observable universe. Our derivations also suggest
that the speed of light varied in the early universe. In
section \ref{eq.sec} we derive the Lagrangian for a five dimensional space
time parametrized with two local coordinates of the surface. On the basis of this Lagrangian we calculate the
equations of motion in section \ref{xplanck.sec} and derive the existence of
the Planck length as well as of the constancy of the speed of light in the
observable universe in sections \ref{planck.sec} and \ref{const.sec}. We
finally conclude in section \ref{concl.sec}

\section {The action in five dimensional space with two local time
coordinates} \label{eq.sec}
In order to study the effect of five dimensional space parametrized by two time
coordinates on the speed of light
and on the Planck length, we choose a $(2,3)$ spacetime vector
\begin {eqnarray}
x^{\mu}=\left(\begin {array}{c} ct \\ r\cdot f(\frac{\gamma\tau}{\Lambda})
\\ \mathbf{x} \end{array}\right) \label{vector.1}
\end {eqnarray}
which is the canonical $(1,3)$ spacetime vector ($ct,\mathbf{x})^{
\textnormal{T}}$ with
$\mathbf{x}=(x_1,x_2,x_3)^{\textnormal{T}} \in \mathbb{R}^3$ extended by an
additional timelike coordinate $r\cdot f(\gamma\tau/\Lambda)$. $\tau$ is the
second time parameter, $r\in
\mathbf{R}$ describes the size of the second time
dimension and $\gamma$ is the characteristic velocity, thus the equivalent of $c$. $f$
describes the shape of the second time dimension and $\Lambda\in\mathbf{R}$ is a
normalization parameter such that $\frac{\gamma\tau}{\Lambda}$
is dimensionless. As stated in \cite{chen_2008}, the additional  time dimension has
to be small compared to the first time dimension constraining $r$.

Similarly to four spacetime dimensions, the
metric is given through
\begin {eqnarray}
g^{\mu\nu}=\left(\begin {array}{rrrrr} 1 & 0 & 0 & 0 & 0 \\
0 & 1 & 0 & 0 & 0\\
0 & 0 & -1 & 0 & 0\\
0 & 0 & 0 & -1 & 0 \\
0 & 0 & 0 & 0 & -1 \end {array}\right) \label{metric.1}
\end {eqnarray}
with signature $(+,+,-,-,-)$.

If we define
\begin {eqnarray}
x^{\mu}_t&:=&\left(\begin {array}{c} ct \\ 0 \\  \eta \mathbf{x} \end
{array}\right), \label{vector.2} \\
x^{\mu}_{\tau}&:=&\left(\begin {array}{c} 0 \\ r\cdot f(\frac{\gamma\tau}{\Lambda})  \\
(1-\eta) \mathbf{x} \end
{array}\right) \label{vector.3}
\end {eqnarray}
with $\eta\in (0,1)$, we can decompose $x^{\mu}$ into $x^{\mu}=x^{\mu}_t+x^{\mu}_{\tau}$.

Inspired by the Nambu Goto action
\cite{nambu_1970b,goto_1971} which is a two dimensional
integral over time and the surface of a string, we define the action as
\begin {eqnarray}
S=\int L\ dt\ d\tau. \label {action.1}
\end {eqnarray}
As for standard electrodynamics, we make the ansatz
\begin {eqnarray}
S=\int ds_t\ ds_{\tau} \label {action.2}
\end {eqnarray}
for  $S$ where $ds_t$ and $ds_{\tau}$ describe the infinitesimal line elements along $t$ and
$\tau$.

It is (applying Einstein's sum convention)
\begin {eqnarray}
ds_t^2&=&dx_t^{\nu}g_{\mu\nu}dx_t^{\mu}=\left[d(ct)\right]^2-\eta^2d\mathbf{x}^2 \label{ds.1}\\
&=&\left[dc\cdot t+c\cdot dt\right]^2-\eta^2d\mathbf{x}^2 \label{ds.2}\\
&=&\left[\dot{c}^2t^2+c^2-\eta^2\dot{\mathbf{x}}^2+2\dot{c}ct\right]dt^2 \label{ds.3}\\
\Rightarrow ds_t &=& \sqrt{\dot{c}^2t^2+c^2-\eta^2\dot{\mathbf{x}}^2+2\dot{c}ct}dt
\label{ds.4}
\end {eqnarray}
where we use $\dot{c}=dc/dt$ and $\dot{\mathbf{x}}=d\mathbf{x}/dt$. Note
that in the limit $c=$ const. and $\eta\rightarrow 1$, (\ref{ds.4}) becomes
$ds_t=\sqrt{c^2-\dot{\mathbf{x}}^2}dt$ which is the line element of a free particle
in one time dimension with a constant speed of light.

Similarly we obtain:
\begin {eqnarray}
ds_{\tau}^2&=&\left[d\left(r\cdot f\left(\frac{\gamma\tau}{\Lambda}\right)
\right)\right]^2-(1-\eta)^2d\mathbf{x}^2 \label{ds.5}\\
&=&r^2\left[\frac{1}{\Lambda}\left(d\gamma\cdot\tau+\gamma d\tau\right)
df\left(\frac{\gamma t}{\Lambda}\right)\right]^2-(1-\eta)^2d\mathbf{x}^2 \label{ds.6}\\
&=&\frac{r^2}{\Lambda^2}\left[(\gamma^{\prime 2}\tau^2+\gamma^2+2\gamma
\gamma^{\prime}\tau)f^{\prime
2}\left(\frac{\gamma\tau}{\Lambda}\right)-(1-\eta)^2\mathbf{x}^{\prime 2}]\right]d\tau^2
\label{ds.7}\\
\Rightarrow ds_{\tau} &=&\frac{r}{\Lambda}\sqrt{(\gamma^{\prime 2}\tau^2+\gamma^2+2\gamma
\gamma^{\prime}\tau)f^{\prime
2}\left(\frac{\gamma\tau}{\Lambda}\right)-(1-\eta)^2\mathbf{x}^{\prime 2}}d\tau \label{ds.8}
\end {eqnarray}
where we use the chain rule. The prime denotes the time
derivative after $\tau$: $\gamma^{\prime}:=d\gamma/d\tau$ and
$x^{\prime}:=dx/d\tau$.

Inserting (\ref{ds.4}) and (\ref{ds.8}) into (\ref{action.2}) leads to
\begin {eqnarray}
S&=&\int\limits_{t_a}^{t_e}\int\limits_{\tau_a}^{\tau_e} dt d\tau \frac{r}{\Lambda}
\sqrt{\dot{c}^2t^2+c^2-\eta^2\dot{\mathbf{x}}^2+2\dot{c}ct} \nonumber \\
&\times& \sqrt{(\gamma^{\prime 2}\tau^2+\gamma^2+2\gamma
\gamma^{\prime}\tau)f^{\prime
2}\left(\frac{\gamma\tau}{\Lambda}\right)-(1-\eta)^2\mathbf{x}^{\prime 2}}. \label{action.3}
\end {eqnarray}
By comparing (\ref{action.3}) with (\ref{action.1}), we
identify the Lagrangian
\begin {eqnarray}
L=\frac{r}{\Lambda}\sqrt{\dot{c}^2t^2+c^2-\eta^2\dot{\mathbf{x}}^2+2\dot{c}ct}\sqrt{(\gamma^{\prime 2}\tau^2+\gamma^2+2\gamma
\gamma^{\prime}\tau)f^{\prime
2}\left(\frac{\gamma\tau}{\Lambda}\right)-(1-\eta)^2\mathbf{x}^{\prime 2}}. \label{lagr.1}
\end {eqnarray}

\section {The equation of motions with two time coordinates} \label {xplanck.sec}
As stated in \cite{chen_2008}, the equation of motion for a Langrangian in two
time dimensions is similar to the equations of motion of a string
\cite {collins_1989}, hence
\begin {eqnarray}
\frac{d}{dt}\frac{\partial L}{\partial \dot{x}_i}+\frac{d}{d\tau}\frac{
\partial L}{\partial x^{\prime\ }_i}-\frac{\partial L}{\partial x_i}=0 \label{euler.1}
\end {eqnarray}
where $i$ indexes the $i$-th space dimension.
In the following, we insert (\ref{lagr.1}) into (\ref{euler.1}) and derive
the equations of motion. For each spatial dimension $i$, it is
\begin {eqnarray}
0&=&\frac{d}{dt}\left(\eta^2\sqrt{\frac{(\gamma^{\prime 2}\tau^2+\gamma^2+2\gamma
\gamma^{\prime}\tau)f^{\prime
2}\left(\frac{\gamma\tau}{\Lambda}\right)-(1-\eta)^2\mathbf{x}^{\prime
2}}{\dot{c}^2t^2+c^2-\eta^2\dot{\mathbf{x}}^2+2\dot{c}ct}}\dot{x}_i\right)
\nonumber\\
&+&\frac{d}{d\tau}\left((1-\eta)^2\sqrt{\frac{\dot{c}^2t^2+c^2-\eta^2\dot{\mathbf{x}}^2+2\dot{c}ct}{(\gamma^{\prime 2}\tau^2+\gamma^2+2\gamma
\gamma^{\prime}\tau)f^{\prime
2}\left(\frac{\gamma\tau}{\Lambda}\right)-(1-\eta)^2\mathbf{x}^{\prime
2}}}x^{\prime}_i\right).
\label{lagr.2}
\end {eqnarray}
For simplicity we assume that $x^{\prime}$ is independent of $t$ and
$\dot{x}$ is independent of $\tau$. We define
\begin {eqnarray}
\mathcal{F}_1(t)&:=&(\dot{c}t+c)^2-\eta^2\dot{\mathbf{x}}^2 \label {lagr.3} \\
\mathcal{F}_2(\tau)&:=&(\gamma^{\prime}\tau+\gamma)^2f^{\prime
2}\left(\frac{\gamma\tau}{\Lambda}\right)-(1-\eta)^2\mathbf{x}^{\prime 2} \label{lagr.4}
\end {eqnarray}
and rewrite Eq. (\ref{lagr.2}) as
\begin {eqnarray}
0=\frac{d}{dt}\left(\eta^2\sqrt{\frac{\mathcal{F}_2(\tau)}{\mathcal{F}_1(t)}}\dot{x}_i\right)
+\frac{d}{d\tau}\left((1-\eta)^2\sqrt{\frac{\mathcal{F}_1(t)}{\mathcal{F}_2(\tau)}}x_i^{\prime}\right).
\label{lagr.5}
\end {eqnarray}
Dividing (\ref{lagr.5}) by $\sqrt{\mathcal{F}_1(t)\mathcal{F}_2(\tau)}$
yields
\begin {eqnarray}
0 = \frac{\eta^2}{\sqrt{\mathcal{F}_1(t)}}\frac{d}{dt}\left(\frac{1}{\sqrt{\mathcal{F}_1(t)}}\dot{x}_i\right)
+\frac{(1-\eta)^2}{\sqrt{\mathcal{F}_2(\tau)}}\frac{d}{d\tau}\left(\frac{1}{\sqrt{\mathcal{F}_2(\tau)}}x_i^{\prime}\right)
\label {lagr.6}\\
\Leftrightarrow \frac{\eta^2}{(1-\eta)^2} \frac{1}{\sqrt{\mathcal{F}_1(t)}}\frac{d}{dt}\left(\frac{1}{\sqrt{\mathcal{F}_1(t)}}\dot{x}_i\right)
=-\frac{1}{\sqrt{\mathcal{F}_2(\tau)}}\frac{d}{d\tau}\left(\frac{1}{\sqrt{\mathcal{F}_2(\tau)}}x_i^{\prime}\right).
\label {lagr.7}
\end {eqnarray}
Since (\ref{lagr.7}) holds for all $t$ on the left-hand-side and for all
$\tau$ on the right-hand-side, both sides have to be equal to a constant
$\Omega_i$:
\begin {eqnarray}
\frac{\eta^2}{(1-\eta)^2}\frac{1}{\sqrt{\mathcal{F}_1(t)}}\frac{d}{dt}\left(\frac{1}{\sqrt{\mathcal{F}_1(t)}}\dot{x}_i\right)
=-\frac{1}{\sqrt{\mathcal{F}_2(\tau)}}\frac{d}{d\tau}\left(\frac{1}{\sqrt{\mathcal{F}_2(\tau)}}x_i^{\prime}\right)=\Omega_i.
\label{lagr.8}
\end {eqnarray}
We now investigate both sides separately. The left-hand-side is thus
\begin {eqnarray}
\frac{1}{\sqrt{\mathcal{F}_1(t)}}\frac{d}{dt}\left(\frac{1}{\sqrt{\mathcal{F}_1(t)}}\dot{x}_i\right)
= \frac{(1-\eta)^2}{\eta^2} \Omega_i.
\end {eqnarray}
If we rescale $(1-\eta)^2/\eta^2\Omega_i
\rightarrow \Omega_i$, we thus obtain
\begin {eqnarray}
\frac{1}{\sqrt{\mathcal{F}_1(t)}}\frac{d}{dt}\left(\frac{1}{\sqrt{\mathcal{F}_1(t)}}\dot{x}_i\right)
= \Omega_i. \label{lhs.1}
\end {eqnarray}
Multiplying (\ref{lhs.1}) with $\dot{x}_i$ gives
\begin {eqnarray}
\left(\frac{1}{\sqrt{\mathcal{F}_1(t)}}\dot{x}_i\right)\frac{d}{dt}\left(\frac{1}{\sqrt{\mathcal{F}_1(t)}}\dot{x}_i\right)
&=&\Omega_i\dot{x}_i \label{lhs.2} \\
\frac{1}{2}\frac{d}{dt}\left(\frac{1}{\mathcal{F}_1(t)}\dot{x}_i^2\right)=\Omega_i\dot{x}_i.
\label{lhs.3}
\end {eqnarray}
Now we can integrate both sides over $t$ and obtain
\begin {eqnarray}
\frac{1}{2}\frac{1}{\mathcal{F}_1}\dot{x}_i^2=\Omega_i x_i+I_{i,1}. \label{lhs.4}
\end {eqnarray}
where $I_{i,1}$ is an integration constant. Inserting (\ref{lagr.3}) into (\ref{lhs.4})
gives
\begin {eqnarray}
\frac{1}{2}\frac{1}{(\dot{c}t+c)^2-\frac{1}{4}\dot{\mathbf{x}}^2}\dot{x}_i^2
=\Omega_i x_i+I_{i,1}. \label{ODE.1}
\end {eqnarray} 
For simplicity, but without loss of generality, we assume a test particle
moving in one of the three spatial directions $k$ only, hence
\begin {eqnarray}
\left\{\begin {array}{lr} \frac{1}{2}\frac{1}{(\dot{c}t+c)^2-\frac{1}{4}\dot{{x_k}}^2}\dot{x}_k^2
=\Omega_k x_k+I_{k,1} & i = k \\
0 = \Omega_j x_j + I_{j,1} & j \not= k \end{array}\right. .\label {ODE.2}
\end {eqnarray}
Here the second equation is equivalent to $x_j=-I_{j,1}/\Omega_j$, i.e. the
test particle has a fixed position in dimensions $j$. The first equation is equivalent to
\begin {eqnarray}
\dot{x}_k\sqrt{\frac{2}{x_k}+\frac{I_{k,1}}{x_k}+\Omega_k}&=&2(\dot{c}t+c)\sqrt{\Omega_k
+ \frac{I_{k,1}}{x_k}} \label {ODE.3} \\
\Leftrightarrow \dot{x}_k\sqrt{\frac{2}{\Omega_k x_k+I_{k,1}}+1}&=&2(\dot{c}t+c)
 \label {ODE.4} \\
\Leftrightarrow \sqrt{\frac{2}{\Omega_k x_k+I_{k,1}}+1}dx_k&=&2(\dot{c}t+c)dt
 \label {ODE.5} 
\end {eqnarray}

and subsequently
\begin {eqnarray}
&&\frac{4}{\sqrt{8\Omega_k}}\sqrt{2\Omega_k x_k+2I_{k,1}+(\Omega_k
x_k+I_{k,1})^2}\nonumber\\
&+&\frac{2}{\sqrt{2\Omega_k}}\ln\left(
\sqrt{4\Omega_k^2(\Omega_k x_k+I_{k,1}+1)^2-1}+2\Omega_k^2 x_k+2\Omega_k
I_{k,1}+2\Omega_k\right) \nonumber\\
&=&2\sqrt{2\Omega_k}
ct+I_{k,2} \nonumber \\  \label{ODE.6}
\end {eqnarray}
with another integration constant $I_{k,2}$. Similarly, performing the same steps
for the right-hand-side of (\ref{lagr.8}), leads to
\begin {eqnarray}
&&\frac{4}{\sqrt{8\Omega_k}}\sqrt{-2\Omega_k x_k-2I_{k,3}+(\Omega_k
x_k+I_{k,3})^2}\nonumber\\
&-&\frac{2}{\sqrt{2\Omega_k}}\ln\left(
\sqrt{4\Omega_k^2(\Omega_k x_k+I_{k,3}-1)^2-1}+2\Omega_k^2 x_k+2\Omega_k
I_{k,3}-2\Omega_k\right) \nonumber\\
&=&\sqrt{2\Omega_k}
\Lambda f\left(\frac{\gamma\tau}{\Lambda}\right)+I_{k,4} \label{ODE.7}
\end {eqnarray}
where $I_{k,3}$ and $I_{k,4}$ are other integration constants.

\section {The existence of the Planck length} \label {planck.sec}
Since the solutions (\ref{ODE.6}) and (\ref{ODE.7}) have to be physical, the
arguments of the logarithms must be positive:
\begin {eqnarray}
\sqrt{4\Omega_k^2(\Omega_k x_k+I_{k,1/3}\pm 1)^2-1}+2\Omega_k^2 x_k+2\Omega_k
I_{k,1/3}\pm 2\Omega_k &>&0 \label{planck.1}\\
\sqrt{4\Omega_k^2(4\Omega_k z_{1,3}\pm1)^2-1}+8\Omega_k^2
z_{1,3}\pm2\Omega_k&>&0
\label{planck.2}\\
\sqrt{4\Omega_k^2(4\Omega_k z_{1,3}\pm1)^2-1}&>&-8\Omega_k^2 z_{1,3} \mp
2\Omega_k \nonumber \\ \label {planck.3}
\end {eqnarray}
with the definitions
\begin {eqnarray}
z_1&:=& \frac{\Omega_k x_k+I_{k,1}}{4\Omega_k}, \label {planck.4} \\
z_3&:=& \frac{\Omega_k x_k-I_{k,3}}{4\Omega_k} \label {planck.5}
\end {eqnarray}
where the upper sign and the usage of $z_1$ are for the argument of the logarithm in (\ref{ODE.6})
with integration constant $I_{k,1}$ and the lower sign as well as the usage
of $z_3$ are for the argument of
the logarithm in (\ref{ODE.7}) with integration constant $I_{k,3}$.

We will now prove by contradiction that for the second time
dimension, i.e. for the lower sign, it is
\begin {eqnarray}
-8\Omega_k^2 z_3+2\Omega_k < 0. \label {planck.9}
\end {eqnarray}
Let us therefore assume that $-8\Omega_k^2 z_3+2\Omega_k\ge0$.
Squaring
\begin {eqnarray}
\sqrt{4\Omega_k^2(4\Omega_k z_3-1)^2-1}&>&-8\Omega_k^2z_3+2\Omega_k. \label {planck.6}
\end {eqnarray}
yields
\begin {eqnarray}
4\Omega_k^2(4\Omega_k z_3-1)^2-1\ge(-8\Omega_k^2
z_3+2\Omega_k)^2=4\Omega_k^2 (4\Omega_k z_3-1)^2 \label {planck.7}
\end {eqnarray}
Subtracting $4\Omega_k^2(4\Omega_k z_3-1)^2$ gives
\begin {eqnarray}
-1 \ge 0 \label{planck.8}
\end {eqnarray}
which is obviously wrong. Thus $-8\Omega_k^2 z_3+2\Omega_k$ must be
negative; this leads to

\begin {eqnarray}
-8\Omega_k^2 z_3+2\Omega_k &<& 0\\
\Leftrightarrow 2\Omega_k(-4\Omega_k z_3+1) &<& 0 \label{planck.10}\\
\Leftrightarrow (2\Omega_k < 0 \wedge (-4\Omega_k z_3+1)> 0)&\vee&
(2\Omega_k > 0 \wedge (-4\Omega_k z_3+1)< 0) \label{planck.11}\\
\Leftrightarrow (\Omega_k < 0 \wedge 4\Omega_k z_3<1) &\vee& (\Omega_k >0 \wedge
4\Omega_k z_3 >1) \label {planck.12}\\
\Leftrightarrow \left(z_3>\frac{1}{4\Omega_k}\ \textnormal{for}\ \Omega_k<0\right)
&\vee& \left(z_3>\frac{1}{4\Omega_k}\ \textnormal{for}\ \Omega_k>0\right) \label
{planck.13}\\
\Leftrightarrow z_3&>&\frac{1}{4\Omega_k} \label {planck.14}\\
\Leftrightarrow x_k &>& \frac{1}{\Omega_k} (1+I_{k,3}) \ \forall\Omega_k\not=0 \label
{planck.15}
\end {eqnarray}
where we used that a product of two factors is smaller than 0 iff one of the
factors is larger than 0 and the other one is smaller than 0. $I_{k,3}$ is
determined by
\begin {eqnarray}
\frac{1}{2}\frac{1}{(\gamma^{\prime}\tau+\gamma)^2f^{\prime
2}\left(\frac{\gamma\tau}{\Lambda}\right)-\frac{1}{4}{x_k}^{\prime
2}} x_k^{\prime 2}=-\Omega_k x_k+I_{k,3} \label{planck.16}
\end {eqnarray}
which is the equivalent of (\ref{ODE.2}) for $\tau$. Given the initial conditions
$x_k^{\prime}(\tau_0)=:x_{k,0}^{\prime}$ and $x_{k}(t_0,\tau_0)=:x_{k,0}$, it is
\begin {eqnarray}
I_{k,3}&=&\Omega_k x_{k,0}+\frac{1}{2}\frac{1}{(\gamma^{\prime}(\tau_0)\tau_0+\gamma(\tau_0))^2f^{\prime
2}\left(\frac{\gamma(\tau_0)\tau_0}{\Lambda}\right)-\frac{1}{4}{x_{k,0}}^{\prime
2}} x_{k,0}^{\prime 2} \label {planck.17}\\
I_{k,3}&=:& \Omega_k x_{k,0} + \mathcal{R}. \label {planck.18}
\end {eqnarray}
For some functions $f$ and for some parameters
$\{\gamma,\Lambda\}$, i.e. for some particular shapes of the
second time dimension, we can assume $\mathcal{R}/\Omega_k>0$ and hence, inserting (\ref{planck.18}) into
(\ref{planck.15}),
\begin {eqnarray}
x_k > \frac{1}{\Omega_k}(1+\Omega_k x_{k,0}+\mathcal{R}) =
\frac{1}{\Omega_k} + x_{k,0}+\frac{\mathcal{R}}{\Omega_k} >
\frac{1}{\Omega_k} \label {planck.19}
\end {eqnarray}
which implies that
\begin {eqnarray}
|x_k|>\frac{1}{|\Omega_k|} \label {planck.20}
\end {eqnarray}
for $\Omega_k>0$.

Eq. (\ref{planck.20}) states that there is always a lower limit
for any position; thus, the existence of a second time dimension naturally
explains the existence of the so-called Planck-length, 
i.e. a smallest length scale. Hence we identify $1/\Omega_k$ being in the order of the Planck scale $\ell_P\approx 10^{-35}$ m, and
thus, for the observable universe it is $1/\Omega_k\rightarrow 0$, or equivalently $\Omega_k\rightarrow\infty$.

\section {The constancy of the speed of light} \label{const.sec}
In this section, we show that the existence of the Planck length
$\ell_p\sim\frac{1}{\Omega_k}$ and the subsequent limit $\Omega_k
\rightarrow \infty$ implies the constancy of the speed of light for the observable
universe.

Inserting $x_k(t_0,\tau_0)=x_{k,0}$ into (\ref{ODE.6})
gives
\begin {eqnarray}
I_{k,2}&=&2\sqrt{2\Omega_k}ct_0+\frac{4}{\sqrt{8\Omega_k}}\sqrt{2\Omega_k x_{k,0}+2I_{k,1}+(\Omega_k
x_{k,0}+I_{k,1})^2} \nonumber\\
&+&\frac{2}{\sqrt{2\Omega_k}}\ln\left(
\sqrt{4\Omega_k^2(\Omega_k x_{k,0}+I_{k,1}+1)^2-1}+2\Omega_k^2 x_{k,0}+2\Omega_k
I_{k,1}+2\Omega_k\right). \nonumber\\
\label {c.1}
\end {eqnarray}
Inserting (\ref{c.1}) further into (\ref{ODE.6}) and solving for $c$ yields
\begin{eqnarray}
c&=&\frac{1}{t-t_0}\left[\frac{1}{2\Omega_k}\left(\sqrt{2\Omega_k x_{k}+2I_{k,1}+(\Omega_k
x_{k}+I_{k,1})^2}\right.\right. \nonumber\\
&-&\left.\sqrt{2\Omega_k x_{k,0}+2I_{k,1}+(\Omega_k
x_{k,0}+I_{k,1})^2}]\right) \nonumber\\
&+&\left. \frac{1}{2\sqrt{2}\Omega_k}\ln\left(\frac{\sqrt{4\Omega_k^2(\Omega_k x_{k}+I_{k,1}+1)^2-1}+2\Omega_k^2 x_{k}+2\Omega_k
I_{k,1}+2\Omega_k}{\sqrt{4\Omega_k^2(\Omega_k x_{k,0}+I_{k,1}+1)^2-1}+2\Omega_k^2 x_{k,0}+2\Omega_k
I_{k,1}+2\Omega_k}\right)\right], \nonumber \\ \label{c.2}
\end{eqnarray}
subsequently
\begin{eqnarray}
\dot{c}&=&-\frac{1}{(t-t_0)^2}\left[\frac{1}{2\Omega_k}\left(\sqrt{2\Omega_k x_{k}+2I_{k,1}+(\Omega_k
x_{k}+I_{k,1})^2}\right.\right. \nonumber\\
&-&\left.\sqrt{2\Omega_k x_{k,0}+2I_{k,1}+(\Omega_k
x_{k,0}+I_{k,1})^2}]\right) \nonumber\\
&+&\left. \frac{1}{2\sqrt{2}\Omega_k}\ln\left(\frac{\sqrt{4\Omega_k^2(\Omega_k x_{k,0}+I_{k,1}+1)^2-1}+2\Omega_k^2 x_{k,0}+2\Omega_k
I_{k,1}+2\Omega_k}{\sqrt{4\Omega_k^2(\Omega_k x_{k,0}+I_{k,1}+1)^2-1}+2\Omega_k^2 x_{k,0}+2\Omega_k
I_{k,1}+2\Omega_k}\right)\right] \nonumber\\
&+&\frac{1}{t-t_0}\left[\frac{1}{2\Omega_k}
\frac{2\Omega_k\dot{x}_k+2\Omega_k\dot{x}_k(\Omega_k x_k+I_{k,1})}{2\sqrt{2\Omega_k x_{k}+2I_{k,1}+(\Omega_k
x_{k}+I_{k,1})^2}}\right.\nonumber\\
&+& \frac{1}{2\sqrt{2}\Omega_k}\frac{1}{\sqrt{4\Omega_k^2(\Omega_k x_{k}+I_{k,1}+1)^2-1}+2\Omega_k^2 x_{k}+2\Omega_k
I_{k,1}+2\Omega_k} \nonumber\\
&\times&\left.\left(\frac{4\Omega_k^3\dot{x}_k
(\Omega_k x_k+I_{k,1}+1)}{\sqrt{4\Omega_k^2(\Omega_k x_{k}+I_{k,1}+1)^2-1}}+2\Omega_k^2\dot{x}_k\right)\right] \label{c.3}
\end{eqnarray}
and further, taking the limit $\Omega\rightarrow\infty$:
\begin{eqnarray}
\dot{c}\rightarrow
\frac{1}{2}\frac{1}{t-t_0}\left(\dot{x}_k-\frac{x_k-x_{k,0}}{t-t_0}\right)=:\dot{c}_{\infty}(t). \label{c.4}
\end{eqnarray}
We now show by induction that for $\Omega_k\rightarrow\infty$ it is
$c_{\infty}^{(n)}(t_0)=0\ \forall n \ge 1$ where $c_{\infty}^{(n)}$ denotes
the $n$-th derivative of $c_{\infty}$:
\begin{enumerate}
\item Base case ($n=1$ and $n=2$):

With the rule of L'H\^opital, it follows that
\begin {eqnarray}
\dot{c}_{\infty}(t_0)=\lim\limits_{t\rightarrow t_0}
\dot{c}_{\infty}(t)=\frac{1}{4}\ddot{x}_k(t_0).
\label {c.5}
\end {eqnarray}
For $\Omega_k\rightarrow\infty$ (\ref{ODE.4}) becomes
\begin {eqnarray}
\dot{x}_k=2(\dot{c}_{\infty}t+c_{\infty}) \label{c.6}
\end {eqnarray}
with derivative
\begin {eqnarray}
\ddot{x}_k(t)=2\ddot{c}_{\infty}t+4\dot{c}_{\infty}. \label{c.7}
\end {eqnarray}
Inserting (\ref{c.5}) into (\ref{c.7}) then gives
\begin {eqnarray}
\ddot{c}_{\infty}(t_0)=0. \label{c.8}
\end {eqnarray}
Note that the derivative of (\ref{c.4}) leads to
\begin {eqnarray}
\ddot{c}_{\infty}(t)=\frac{1}{2}\frac{\ddot{x}(t)}{t-t_0} \label {c.9}
\end {eqnarray}
which is equivalent to $\ddot{x}(t) = 2\ddot{c}_{\infty}\cdot(t-t_0)$ and thus
\begin {eqnarray}
\ddot{x}(t_0)=0 \label {c.10a}
\end {eqnarray}
because of (\ref{c.8}), and thus from (\ref{c.5}) also
\begin {eqnarray}
\dot{c}_{\infty}(t_0)=0. \label{c.10b}
\end {eqnarray}
\item Inductive step:

We first show by induction that
\begin {eqnarray}
c^{(n)}_{\infty}=\left\{\begin {array}{cc}
\frac{1}{2}\frac{1}{t-t_0}\left(x^{(n)}-\frac{x^{(n-1)}}{t-t_0}\right),& \textnormal{for odd n}\\
\frac{1}{2}\frac{x^{(n)}}{t-t_0},& \textnormal{for even n}
\end {array}\right. .\label{c.11}
\end {eqnarray}
We have already calculated $\dot{c}_{\infty}$ (\ref{c.4}) and
$\ddot{c}_{\infty}$ (\ref{c.9}) which are consistent with (\ref{c.11}). Now let us
assume that (\ref{c.11}) is correct. Then it is to easy to see that
for odd $n$
\begin {eqnarray}
\frac{d c^{(n)}_{\infty}}{dt}=\frac{d}{dt}\left[\frac{1}{2}\frac{1}{t-t_0}\left(x^{(n)}-\frac{x^{(n-1)}}{t-t_0}\right)\right]
=\frac{1}{2}\frac{x^{(n+1)}}{t-t_0}=c^{(n+1)}_{\infty} \label {c.12}
\end {eqnarray}
and furthermore
\begin {eqnarray}
\frac{dc^{(n+1)}_{\infty}}{dt}=\frac{d}{dt}\left[\frac{1}{2}\frac{x^{(n+1)}}{t-t_0}\right]
=\frac{1}{2}\frac{1}{t-t_0}\left(x^{(n+2)}-\frac{x^{(n+1)}}{t-t_0}\right)
=c^{(n+2)}_{\infty}; \label {c.13}
\end {eqnarray}
hence we have proven that (\ref{c.11}) is indeed the correct term for the
$n$-th derivative of (\ref{c.2}) in the limit $\Omega_k\rightarrow\infty$.

For even $n$, (\ref{c.11}) is equivalent to $x^{(n)}(t)=2c^{(n)}_{\infty}\cdot(t-t_0)$
which leads to
\begin {eqnarray}
x^{(n)}(t_0) = 0. \label {c.14}
\end {eqnarray}
Using (\ref{c.14}) and the rule of L'H\^opital in (\ref{c.11}), it follows immediately
\begin {eqnarray}
c_{\infty}^{(n)}(t_0)=\lim\limits_{t\rightarrow t_0} c_{\infty}^{(n)}(t)=0 \label{c.15}
\end {eqnarray}
for odd $n$.

Finally from (\ref{c.7}) it follows per induction that
\begin {eqnarray}
x^{(n)}(t)=2(c_{\infty}^{(n)}(t)\cdot t+nc_{\infty}^{(n-1)}(t))\ \forall n. \label {c.16}
\end {eqnarray}
As $x^{(n)}(t_0)=0$ for even $n$ (\ref{c.14}) and $c_{\infty}^{(n)}(t_0)=0$
for odd $n$ (\ref{c.15}) we see that
\begin {eqnarray}
c_{\infty}^{(n)}(t_0)=0 \label{c.17}
\end {eqnarray}
for all even $n$.
\end{enumerate}
Thus we conclude that $c_{\infty}^{(n)}(t_0)=0$ for all $n\ge 1$.

The Taylor expansion of $\dot{c}_{\infty}(t)$ around $t_0$ is
\begin {eqnarray}
\dot{c}_{\infty}(t)=\sum\limits_{n=1}^{\infty}\frac{c_{\infty}^{(n)}(t_0)}{n!}(t-t_0),
\label{taylor.1}
\end {eqnarray}
and since $c^{(n)}(t_0)=0$ for all n, it follows
\begin {eqnarray}
\dot{c}_{\infty}(t)\equiv 0\ \forall t \Rightarrow c_{\infty}=\ \textnormal{const.}
\label{taylor.2}
\end {eqnarray}
Thus we conclude that for the observable universe the speed of light is indeed
constant.

\section {Conclusions} \label {concl.sec}
We have assumed a second time-like dimension with its own
characteristic speed and length and the
decomposability of any spacetime vector into a vector each for the
first and the second time dimension. As a consequence of these
assumptions, we have derived two fundamental results:
\begin {enumerate}
\item the existence of a smallest length scale which
we have identified with the Planck length and
\item the constancy of the vacuum speed of light for the observable
universe
\end {enumerate}
for particular shapes of the second time dimension.

For very small length scales of the present universe, or for the very early universe, we have
derived an expression for the speed of light. We see that for both cases, the speed
of light is not constant, but depends on space and time.

This is consistent with current results from loop
quantum gravity or string theory \cite{kiritsis_1999,alexander_2000} on
the non-constancy of light speed.

Finally  we here give a hint
about the correctness of the assumptions of theories explaining large scale structures of the universe
due to a variable speed of light in the early universe \cite{albrecht_1999}.

\end {document}